\documentclass[a4paper,twoside]{article}

\usepackage{epsfig}
\usepackage{subcaption}
\usepackage{calc}
\usepackage{amssymb}
\usepackage{amstext}
\usepackage{amsmath}
\usepackage{amsthm}
\usepackage{multicol}
\usepackage{pslatex}
\usepackage{apalike}
\let\bibhang\relax
\usepackage{natbib}
\usepackage[hidelinks]{hyperref}
\usepackage{hyperref} 
\usepackage{amssymb} 
\usepackage{amsmath,amsfonts}
\usepackage{xspace}
\usepackage{color}
\usepackage{xcolor}
\usepackage{fancybox}
\usepackage{fancyhdr}
\usepackage{ifthen} 
\usepackage{paralist} 

\usepackage[T1]{fontenc}
\usepackage{multirow}
\usepackage{hhline}

\usepackage{array}
\newcolumntype{L}[1]{>{\raggedright\let\newline\\\arraybackslash\hspace{0pt}}m{#1}}
\newcolumntype{C}[1]{>{\centering\let\newline\\\arraybackslash\hspace{0pt}}m{#1}}
\newcolumntype{R}[1]{>{\raggedleft\let\newline\\\arraybackslash\hspace{0pt}}m{#1}}

\usepackage{balance} 
\usepackage{adjustbox}

\usepackage{algorithmic}
\usepackage{textcomp}
\usepackage{booktabs}
\usepackage{microtype}
\usepackage{threeparttable}
\def\BibTeX{{\rm B\kern-.05em{\sc i\kern-.025em b}\kern-.08em
    T\kern-.1667em\lower.7ex\hbox{E}\kern-.125emX}}
\usepackage{tcolorbox}

\usepackage{xcolor}

\newcommand{\SQOne}{SQ1-Strategy}
\newcommand{\SQTwo}{SQ2-Resources}
\newcommand{\SQThree}{SQ3-Roles}
\newcommand{\SQFour}{SQ4-Knowledge}
\newcommand{\SQFive}{SQ5-Competence}
\newcommand{\SQSix}{SQ6-Tools}
\newcommand{\SQSeven}{SQ7-Test environment}
\newcommand{\SQEight}{SQ8-Guidelines}
\newcommand{\SQNine}{SQ9-Prioritization}
\newcommand{\SQTen}{SQ10-Test results}
\newcommand{\SQEleven}{SQ11-Process}
\newcommand{\SQTwelve}{SQ12-SUT}
\newcommand{\SQThirteen}{SQ13-Metrics}
\newcommand{\SQFourteen}{SQ14-Testware}
\newcommand{\SQFifteen}{SQ15-Efficient\&Effective}
\newcommand{\SQSixteen}{SQ16-Satisfaction}

\usepackage{SCITEPRESS}     

\begin{document}

\title{ Software Test Automation Maturity - A Survey of the State of the Practice}

\author{\authorname{Yuqing Wang\sup{1},  Mika V. M{\"a}ntyl{\"a}\sup{1}, Serge Demeyer\sup{2}, Kristian Wiklund\sup{3}, Sigrid Eldh\sup{3}, Tatu Kairi\sup{4} }
\affiliation{\sup{1}M3S research unit, University of Oulu, Oulu, Finland}
\affiliation{\sup{2}Universiteit Antwerpen \& Flanders Make, Antwerp, Belgium}
\affiliation{\sup{3}Ericsson AB,Stockholm, Sweden}
\affiliation{\sup{4}Eficode,Helsinki, Finland}
\email{}
}



\abstract{The software industry has seen an increasing interest in test automation. In this paper, we present a test automation maturity survey serving as a self-assessment for practitioners. Based on responses of 151 practitioners coming from above 101 organizations in 25 countries, we make observations regarding the state of the practice of test automation maturity: a) The level of test automation maturity in different organizations is differentiated by the practices they adopt; b) Practitioner reported the quite diverse situation with respect to different practices, e.g.,  85\% practitioners agreed that their test teams have enough test automation expertise and skills, while 47\% of practitioners admitted that there is lack of guidelines on designing and executing automated tests;  c) Some practices are strongly correlated and/or closely clustered; d) The percentage of automated test cases and the use of Agile and/or DevOps development models are good indicators for a higher test automation maturity level; (e) The roles of practitioners may affect response variation, e.g., QA engineers give the most optimistic answers, consultants give the  most pessimistic answers. Our results give an insight into present test automation processes and practices and indicate chances for further improvement in the present industry.}

\onecolumn \maketitle \normalsize \setcounter{footnote}{0} \vfill

\section{\uppercase{Introduction}
\footnotetext{This is the author's version of the work. It is posted here for your personal use. The definitive version was published in Proceedings of the 15th International Conference on Software Technologies (ICSOFT 2020)}}
\label{sec:introduction}

\noindent With the rise of agile development and the adoption of continuous integration, the software industry has seen an increasing interest in test automation~\citep{CTR2019}. Many organizations invest in test automation but fail to reap the expected benefits~\citep{garousi2016}.
Several process-related problems aggravate these failures: unrealistic goals, lack of management support, the undefined test approach, shortage of expertise and resources, and so forth~\citep{rafi2012,sauer2016}. Prior researchers refer to these problems as lack of test automation maturity which obstructs a true continuous improvement paradigm~\citep{kasurinen2010}.

Despite the research efforts on test automation maturity are increasing, there is little known about the state of practice of test automation maturity in the present industry~\citep{rodrigues2016,wang2018test}. It is important to get an understanding of the current test automation processes and practices by empirical studies in order to observe the  opportunities for continuous improvement and expand the contribution and impact of test automation maturity research~\citep{Eldh2014taim,Garousi2020}.

\newcommand{\RQoneTitle}{RQ1 -- Process maturity}
\newcommand{\RQone}{How mature are current test automaton
processes based on adopted practices?}

\newcommand{\RQtwoTitle}{RQ2 -- Practice maturity}
\newcommand{\RQtwo}{Which practices are more mature and immature in the industry?}

\newcommand{\RQthreeTitle}{RQ3 -- Practice correlation}
\newcommand{\RQthree}{Is there a correlation between the adopted test automation practices?}

\newcommand{\RQfourTitle}{RQ4 -- Organizational factors}
\newcommand{\RQfour}{What are the characteristics of organizational units related to test automation maturity? }

\newcommand{\RQfiveTitle}{RQ5 -- Response variation}
\newcommand{\RQfive}{How do the current roles of practitioners (respondents) affect the response variation?}

In this paper, we explore the state of practice of test automation maturity regarding the adoption of test automation practices. This survey study intends to answer the following research questions: 
\begin{itemize}
    \item \textbf{\RQoneTitle:} \textit{\RQone}
    \item \textbf{\RQtwoTitle:} \textit{\RQtwo}
    \item \textbf{\RQthreeTitle:} \textit{\RQthree}
    \item \textbf{\RQfourTitle:} \textit{\RQfour}
    \item \textbf{\RQfiveTitle:} \textit{\RQfive}
    \end{itemize}

The survey itself is based on an extensive literature survey and validated with four industry experts. We distributed the survey through a variety of channels and received 151 responses from more than 101 organizations located in 25 different countries, most of them from Finland, Belgium and Netherlands. The detailed survey results can be found in the rest of this paper.
 
The remainder of this paper is organized as follows: Section~\ref{sec:related_work} presents the related work. Section~\ref{sec:research_method} describes the research method. Section~\ref{sec:results} presents survey results. Section~\ref{sec:discussion} discusses the implication of survey results. Section~\ref{sec:threats} states threats to validity. Section~\ref{sec:conclusion} concludes this paper and states the future work.

\section{\uppercase{Related work}}
\label{sec:related_work}

 
 \noindent Test automation is an important part of software development process~\citep{pocatilu2002}. It consists of a set of practices conducted in key process areas (KPAs), e.g., test design, test execution, test reporting~\citep{Skoomen2013tmap}. Test automation maturity refers to the level of maturity of a test automation process. Achieving the high level of test automation maturity requires that mature practices are performed in each KPA~\citep{Mitchel1994}. Several examples of the negative effects of immature practices have been reported in the literature. Selecting the wrong test tools for the problems at hand may impact the organizational performance~\citep{garousi2017software}. Automating test cases better suited for manual testing may waste time and money~\citep{garousi2017software}. Defining inappropriate metrics is likely to guide your test automation in the wrong direction~\citep{walia2017}. 


%
%
 
 
 In prior research,  ~\cite{wang2019} studied 25 sources (including 18 test maturity models, 3 journal papers, and 4 recent software testing related technical reports) that covering test automation maturity topics. Referring to their research results, based on those 25 sources, an organization should focus on 13 KPAs to conduct test automation practices. The mature practices must be performed in each KPA were synthesized, see Table~\ref{tab:KPAsofTA}. Those mature practices have been proven to be valid in the present industry by test automation experts from both academia and industry.  
 


\begin{table}[ht]
\small
    \caption{Best practices in KPAs~\citep{wang2019}}
    \footnotesize
    \label{tab:KPAsofTA}
    \begin{tabular}{ l p{1.99cm}  p{3.5cm} } 
    \toprule
        &  KPA & Mature practice\\
	\midrule
       K01 & Test automation strategy & Set explicit strategic plans for test automation \\ \hline
       
       K02 & Resources & Provide enough resources (e.g., skilled people, costs, efforts) for test automation.\\\hline
       
       K03 & Test \newline organization & Assemble and collaborate skilled test staffs perform test automation practices.\\\hline
       K04 & Knowledge \newline transfer & Collect and share test automaton related knowledge\\ \hline
       K05 & Test tools & Use right test tools to support testing activities.\\\hline
       K06 & Test \newline environment & Set up test environment with required software, hardware, test data, etc., to execute automated tests\\\hline
       K07 &Test \newline requirements & Derive explicit test automation requirements. \\\hline
       K08 & Test design & Use the specific test design techniques to create maintainable test cases for test automation.\\\hline
       K09 &Test execution & Prioritize and execute automated test cases\\\hline
       K10 &Verdicts & Collect and report useful results of executing automated test cases.\\\hline
       K11 &Test automation \newline process & Define the specific procedures to conduct test automation. \\\hline
       K12 &Measurements & Use right metrics to measure the quality and performance of test automaton.\\ \hline
       K13 & Software Under Test (SUT) & Prepare the testable SUT to be automatically tested. \\
       \bottomrule
    \end{tabular}
    
\end{table}


\section{\uppercase{Research method}}
\label{sec:research_method}
\noindent To answer research questions of this study, we conducted a test automation maturity survey in the industry. The survey was designed and executed following the `Guidelines for Conducting surveys in Software Engineering' from ~\cite{linaaker2015guidelines}. We designed and conducted our survey in six stages: (1) sampling  and distribution plan, (2) survey design, (3) survey validity, (4) survey execution and data collection, (5) measurement of response quality, (6) data analysis. Each stage is described in the following sub-sections. 

\paragraph{ (1) Sampling and distribution plan.}
The first author defined a sampling and distribution plan that specifies the process to select a sample; the second and third authors reviewed this plan. We planned to reach respondents based on their interests in the survey and accessibility using the Convenience Sampling method~\citep{etikan2016}.
This is the dominant approach widely used in survey and empirical research in SE~\citep{sjoberg2005}.
The target population was identified as practitioners directly working on test automation in the whole industry. Social media's interest groups as well as the network of industry contacts were planned to be used to reach the target population and distribute the survey.


\paragraph{ (2) Survey design.}
The survey in this study was administered using an online survey tool Limesurvey. The survey consists of three main parts. Part 1 introduces the survey, including details about the content, a consent statement, steps to answer the survey, and the names of the principal researchers.

\begin{table*}[ht]
  \begin{center}
   \footnotesize
    \caption{Test automation maturity questions in the Part 2}
    \label{tab:SQinPart3}
    \begin{tabular}{p{13.8cm} l}
     \toprule
       Survey questions &  KPA(s)\\ 
      \midrule
      \textbf{\SQOne}. We have a test automation strategy that defines `what test scope will be automated to what degree, when, by whom, by which methods, by what test tools, in what kind of environment'. & K01  \\ \hline
      \textbf{\SQTwo}. We allocate enough resources for test automation, e.g., skilled people, the funding, the time \& effort, test environment with the required software, hardware, or test data for test automation. & K02 \\\hline
      \textbf{\SQThree}. We clearly define roles and responsibilities of stakeholders in test automation. &  K03 \\\hline
      \textbf{\SQFour}. We are systematically learning from prior projects. We collect and share expertise, good test automation practices, and good test tools for future projects.  & K04\\\hline
     \textbf{\SQFive}. Our test team has enough expertise and technical skills to build test automation based on our requirements. & K03 \\\hline
      \textbf{\SQSix}. We currently have the right test tools that best suit our needs. & K05 \\\hline
      \textbf{\SQSeven}. We have control over the configuration of our test environment. & K06 \\\hline
      \textbf{\SQEight}. We have guidelines on designing and executing automated tests. Those guidelines include, e.g., coding standards, test-data handling methods, specific test design techniques to create test cases, processes for reporting and storing test results, the general rules for test tool usage, or information on how to access external resources.& K08, \newline K09 \\\hline
      \textbf{\SQNine}. We effectively prioritize and schedule automated test cases for the execution. & K09 \\\hline
      \textbf{\SQTen}. We are capable to manage and integrate test results collected from different sources (e.g., different test tools, test levels, test phases) into a big picture, and then report useful information to the relevant stakeholders & K10 \\\hline
      \textbf{\SQEleven}. We organize our test automation activities in the stable and controllable test process.& K11\\\hline
      \textbf{\SQTwelve}. Our Software Under Test enables us to conduct our test automation, e.g., maturity, running speed, or testability of our Software Under Test is not a problem for our test automation.	& K13\\\hline
      \textbf{\SQThirteen}. We have the right metrics to measure and improve our test automation process. & K12\\\hline
      \textbf{\SQFourteen}. Our testware (e.g., test cases, test data, test results, test reports, expected outcomes, and other artifacts generated for automated tests) is well organized in a good architecture and it is easy to be maintained. & K08\\\hline
     \textbf{\SQFifteen}. We create automated tests that are able to produce accurate and reliable results in timely fashion. & K09\\\hline
     \textbf{\SQSixteen}. We create automated tests can meet the given test purposes and consequently bring substantial benefits for us, e.g., better detection of defects, increase test coverage, reduce test cycles, good Return on Investment, better guarantee product quality. & K07\\
\bottomrule
    \end{tabular}
  \end{center}

\end{table*}

Part 2 contains test automation maturity questions~(Table~\ref{tab:SQinPart3}). Those questions were defined according to mature practices that must be performed in KPAs as presented in Table~\ref{tab:KPAsofTA}. Each question addresses at least one KPA, see Table~\ref{tab:SQinPart3}. As the questions were revised according to the feedback of industry experts (see Section~\ref{sec:surveyValidity}), there is no one-to-one mapping between questions and KPAs. The purpose of asking those questions is to examine the maturity state of practitioners' test automation processes by checking if they are performing mature practices in the present industry. To answer the question, respondents were expected to indicate their degree of agreement with the statement of each question by using a six-point scale: 1- strongly disagree,  2 - disagree, 3 - slightly disagree, 4 - slightly agree, 5 - agree, 6 - strongly agree. In addition to the scale, we also offered a `no answer' option. The higher point marked in the scale means more mature practices were performed. If a respondent marks a higher
point for all questions, it would correspond with a
higher level of test automation maturity. Part 2 of the survey finished with a free-text field to gather more detailed insights into responses.

Part 3 presents background questions to collect demographic information of respondents and their organizational units to conduct test automation. \cite{kitchenham2002} recommended putting such questions at the end in order to ensure potential respondents will not be deterred from completing the survey.



To boost the response rate, we implemented a reward mechanism. We sent an individual report with a snapshot of the results and a comparison against a baseline of average responses. \cite{ghazi2018survey} reported that software practitioners may not want to participate in a survey that collects information about respondents. Therefore, we designed the survey to be anonymous, both in collecting the responses and reporting the survey results. Also, the survey adheres to the European General Data Protection Regulation (GDPR).

\paragraph{ (3) Survey validity.}
\label{sec:surveyValidity}
We conducted an expert review and a small-scale pilot to evaluate the validity of our survey, in order to examine whether it measures what was intended and avoid potential measurement errors~\citep{litwin1995}.

 Four industrial experts were selected from our network to participate in the review. They come from three Swedish and Finnish companies.
The important criterion to select industrial experts was "hands-on experience with test automation".
The selected experts have been working on test automation for decades.
We allowed all experts to assess a survey sample via the Limesurvey system.
They were asked to review all survey questions and give suggestions for revision.
Several online meetings  were  conducted. As the result, the following revisions were made to the survey: a) revise survey questions; b) make survey questions to be close to everyday situations in an industrial context; c) add concrete industrial examples to explain the practices.

After the expert review, we piloted our survey with three practitioners. They work in the same test automation team; each answered the survey independently. We compared the responses of three practitioners, and found that the differences are rather small which confirmed that the questions were understandable.

\paragraph{ (4) Survey execution and data collection.}
The survey was activated in the Limesurvey system in December 2018, and remained open until June 2019. Survey researchers recommend to ensure the longer duration of availability of online surveys, in order to boost the response rate~\citep{nulty2008}. 

 The survey was distributed to the target population by different methods. First, we posted the survey in software testing related groups in social media: LinkedIn, Facebook, Twitter, and Reddit. Second, we used the email list of TESTOMAT project and FiSTB (mailing list of software testing practitioners in Finland) to distribute the  survey. Within TESTOMAT project, we requested at least one person from each partner company to answer our survey. Third, nine principal contact people from our industry network further distributed our survey in their networks. They sent the survey to potential respondents in their industry contacts via individual emails. 
 
 Table~\ref{tab:statistics_composition_dataset} shows the statistics about data collection in different distribution channels. In this table, `Reached audience' refers to the number of people who view our posts or emails. `Interested respondents' indicate the number of people visited our survey system. `Received responses' indicate the number of people who actually answered the survey. 

 \begin{table}[ht]
 \small
 \begin{threeparttable}
    \caption{Statistics on data collection}
    \label{tab:statistics_composition_dataset}
    \begin{tabular}{p{1.9cm}  p{1.33cm}  p{1.3cm}  l } 
    \hline
      Channels & Reached & Interested &Received \\
 &  audience & audience & responses\\
      \hline
      Social media:&  &  &  \\
       LinkedIn & $>$330* & 81 & 32  \\
       Facebook & $>$617** & 9 & 5   \\
       Twitter & 2250 & 19 & 4   \\
       Reddit & 75& 10 & 2   \\\hline
       Email lists: &  &  &  \\
       TESTOMAT & 89 & 83 & 31  \\
       FiSTB & 920 & 41 & 14  \\ \hline
       Individual & 113  & 91 & 46 \\
       emails &  & & \\\hline
     Total  & $>$4394 & 334 & 157 \\
     \hline
    \end{tabular}
    \begin{tablenotes}
    \scriptsize \item [*]{ LinkedIn only allows to track post views for individual accounts. Our original post had 330 views and it was shared by 5 reposts.}   
     \scriptsize \item [**]  { Facebook does not track post views for individual accounts. 617 is the total member count for all test automation interest groups where we posted the survey. } 
   \end{tablenotes}
    \end{threeparttable}
\end{table}

\paragraph{ (5) Measurement of response quality.}
  The quality of responses in an online survey is rather difficult to control due to its openness~\citep{nulty2008}. In this survey, we used the standard to measure response quality for online surveys from \cite{ganassali2008}. We collected the following measures to assess the quality of 157 responses in our initial pool:  
  
   \begin{itemize}
  \item The response rate: the number of received responses divided by the number of interested respondents (see Table~\ref{tab:statistics_composition_dataset}). 
    \item The dropout rate: the ratio of respondents who started the survey but never reached the end.
    \item The ``same response'' rate: the ratio of respondents who have selected the same response from 6 point scale for all questions in part 2 of the survey. 
    \item The ``no answers'' rate: the ratio of respondents having more than 50\% of 'no answers' on questions in part 2 of survey. 
\end{itemize}

Table~\ref{tab:responsequality} shows the measurement results. To improve the overall response quality, we removed 6 responses we deemed inadequate based on the quality measurements listed above. Thus we removed responses that (a) failed to reach the end of the survey; (b) had  the same response for all questions in part 2; (c) had more than 50\% `no answers'. At the end, a final pool of 151 responses was collected. 

  \begin{table}[ht]
  \small
  \begin{center}
    \caption{Measures for the response quality}
    \label{tab:responsequality}
    \begin{tabular}{p{4cm}  R{1.8cm} } 
    \hline
      The response rate:   & 47.0\%\\
      The dropout rate:   & 5.1\%\\
      The same response rate:   & 1.9\%\\
      The `no answers' rate: & 1.9\% \\
     \hline
    \end{tabular}
  \end{center}
\end{table}
\vspace{-4pt}

\paragraph{ (6) Data analysis.}
\label{ref:dataAnalysis}

The survey data was exported from the Limesurvey system and imported into R for data analysis. Responses on the 6-point scale were recorded in a score 1-6; `no answer' converts to `NA'. The total score of all questions of each response was recorded as maturity score. The data analysis methods for certain research questions are described as follows:


\textit{a. NA omit.} In `\RQoneTitle', we omitted 24 responses having NA value(s) and only analyzed the remaining 127. Since NA has no numeric value, it may affect the calculation of final maturity score (that indicates the level of test automation maturity) of a response. 

\textit{b. Correlation coefficient.} To answer `\RQthreeTitle', the Kendall Rank (KR) correlation coefficient ($r_k$)  was computed between the responses to each test automation maturity question and another in part 2 of our survey. KR correlation coefficient method was finally selected, as our variables (survey questions) are measured on an ordinal scale and follow a monotonic relationship~\citep{puth2015}. 
In this study, R package `col' was used to compute $r_k$, and NA values were automatically excluded in the computing operations by using R function `na.exclude()'. Cohen's~\cite{cohen1988} interpretation is used to describe the strength of the correlation based on the absolute value of $r_k$: weak correlation ($r_k$=.10-.29), medium correlation ($r_k$=.30-.49); strong correlation ($r_k$>.50). Cluster analysis was performed to identify multiple types of practices that fall into relative clusters. R package `hclust' was used to plot the hierarchihcal clustering based on $r_k$-based distance.



\textit{c. Negative binomial regression analysis.}
\label{sec:nbanalysis}
To answer `\RQfourTitle' and `\RQfiveTitle', negative binomial (NB) regression analysis was carried out. We selected this regression model for
three reasons: a) regression analysis is statistical method to examine the relationship among variables; b) the values of our dependent variable are non-negative integers, making NB regression better choice than normal linear regression; c)~NB regression allows for more variability and dispersion compared to linear regression. 

In `\RQfourTitle', we built the NB regression model with characteristics of organizational units as independent variables and respondents' maturity score as the dependent variable. Thus, the independent variables include: the organizational level, the size, who performs (test automation), percentage of automated test cases, and software development lifecycle (SDL). The SDL related variables were created according to three groups of models: agile, traditional (Waterfall and Rational Unified Process), and DevOps. This SDL classification was introduced by~\cite{noll2019agile}. In our classification, DevOps is not combined with agile and it was created as an independent category, since it extends agile principles~\citep{virmani2015}. 

In `\RQfiveTitle', we built the NB regression model with the roles of respondents as independent variables and respondent's maturity score as the dependent variable. The roles of respondents related variables include job positions and work areas. 
 
 The categorical variables were transformed into dummy variables. We used R function `glm.nb()' to build NB regression models and NA values were automatically excluded in computing operations by using R function `na.exclude()'. The analysis on the models was performed to answer research questions. 


\section{Results}
\label{sec:results}
\noindent To present the survey results, we first provide an overview of the respondent profile and organizational unit profile. Next, we answer each research question in turn.

\paragraph{Respondent profile.}
Based on our statistics, 151 respondents of our survey come from more than 101 organizations in 25 different countries. Since our survey was anonymous, we only counted the number of organizations that respondents voluntarily provided. Table~\ref{tab:countryOfRespondent} shows the classification of respondents by country. It is noted that approximately 84.7\% of respondents (N=128) come from Europe. Respondents coming from Finland (27.8\%), Belgium (16.6\%), and Netherlands (13.2\%) made a substantial contribution. The rest of respondents scattered over other countries in the world.  Table~\ref{tab:domain} shows the classification of respondents by sector. Based on that, we see that test automation has been widely adopted in a variety of sectors. 

\begin{table}[ht]
  \begin{center}
  \small
    \caption{Respondents by country}
    \label{tab:countryOfRespondent}
    \begin{tabular}{ p{1.74cm} p{1.55cm} p{1.32cm} p{1.2cm} } 
    \toprule
      {Country} & 
       {N (\%)}&  {Country} & 
       {N (\%)}    \\
      \midrule
      
      Europe: &  &  Asia: & \\
      Finland  & 42 (27.8 \%) & China & 6 (4.0\%) \\
      Sweden & 11 (7.3\%)& Vietnam & 2 (1.3\%) \\
      
      Belgium & 25 (16.6\%)& India & 3 (2.0\%) \\ 
      Netherlands& 20 (13.2\%) & Americas: & \\
      German & 7 (4.6\%) & USA & 3 (2.0\%) \\

      UK & 3 (2.0 \%) & Canada & 3 (2.0\%)  \\
      Spain & 9 (6.0\%)& Others: & 6 (4\%)\\
    
      Turkey & 4 (2.6\%) && \\
      
      Other Europe & 7 (4.6 \%)&& \\ \hline
      
    \end{tabular}
  \end{center}
\end{table}

  \begin{table}[ht]
  \begin{center}
  \small
    \begin{threeparttable}
    \caption{Respondents by sector}
    \label{tab:domain}
    \begin{tabular}{l l} 
    \toprule
      Sector & 
\small{Responses}    \\
      \hline
      Automotive  & 10 ( 6.6 \% )\\
      Transportation \& logistics & 18  (11.9 \%) \\
      Energy and utilities & 11 (7.3\%9 \\
      Financial services & 20 (13.2\%) \\
      Healthcare and life sciences & 16 ( 8.8 \%) \\
      Government and public sector & 13 (8.6 \%) \\
      Telecommunication & 7 (4.6\%)\\
      Software & 94 (62.3\%) \\
      Data processing and hosting & 15 ( 9.9\%) \\
      Industrials & 10 (6.6\%) \\
      Technology hardware and equipment & 9 (6\%) \\
      Others & 27 (17.9\%) \\
      \hline
    \end{tabular}
    \begin{tablenotes}
      \scriptsize \item [*]{ Note that a respondent can work in multiple sectors.}   \end{tablenotes}
     \end{threeparttable}
  \end{center}
\end{table}


Table~\ref{tab:respondentPositions} shows the classification of respondents by job position and work area. Over half of the respondents (56.2\%) have the technical role of testers, QA engineers, and developers. A smaller yet substantial proportion  of respondents (33.1\%) having the managerial role of test leads, managers, and directors. The 10.6 \% of respondents have other roles like consultant and environment architect in test automation.  Respondents are working in a variety of areas in test automation.

  \begin{table}[ht]
  \small
  \begin{center}
  \begin{threeparttable}
    \caption{The current roles of respondents}
    \label{tab:respondentPositions}
    \begin{tabular}{p{3.5cm} p{1.3cm} p{1cm}  } 
    \toprule
      \textbf{} &\small{ Responses} & \%  \\
      \hline
      \textbf{Job positions:} & & \\
      Test Lead/Manager/Director  & 50 & 33.1\% \\
      Tester  & 28& 18.5\% \\
      QA Engineer & 28& 18.5\% \\
      Developer &29 & 19.2\% \\ 
      Consultant &9 & 6.0\% \\ 
      Other & 7 & 4.6\% \\ \hline
      
      \textbf{Work areas*:} & & \\
      Test management  & 76 & 50.3\% \\
      Test tool selection  & 64 & 42.4\% \\
      Test tool usage & 86& 57.0\% \\
      Test design &83 & 55.0\% \\ 
      Test execution &70 & 46.4\% \\ 
      Test environment & 89 & 58.9\% \\
      Test requirements & 91 & 60.3\% \\
      Measurements &64 & 42.4\% \\
      Others & 12& 7.9\% \\
     \bottomrule 
    \end{tabular}
        \begin{tablenotes}
     \scriptsize \item [*]{ Note that a respondent can work in several areas.}   
   \end{tablenotes}
    \end{threeparttable}
  \end{center}
\end{table}

\paragraph{Organizational unit profile.} 
\label{sec:organizationProfile}
Table~\ref{tab:organizationalUnitProfile} shows the organizational unit profile of respondents, i.e. the unit that conducts test automation.
We see that respondents' test automation processes are mainly organized at the team level (34.4\%), the project level (30.5\%), and equally important the whole organization level (32.5\%). Only few are organized at another level (2.6\%). The organizational units differ in size; we have more responses from smaller organizational units. 

Respondents reported that their test automation processes are mainly conducted by in-house test team (62.9\%) and developers~(50.3\%) in their organizational units. In only 19.2\% of the cases, test automation is outsourced.

  \begin{table}
  \small
  \begin{center}
      \begin{threeparttable}
    \caption{Organizational unit profile}
    \label{tab:organizationalUnitProfile}
    \begin{tabular}{l l} 
    \toprule
       & \textbf{Responses}  \\
      \hline
      \textbf{The level:}  & \\
      The team level  & 52  (34.4\%) \\
      The project level  & 46  (30.5\%) \\
      The organization level &49  (32.5\%) \\
      Other level & 4 (2.6\%) \\ \hline
      
      \textbf{The size:} &  \\
      Micro size (1-10 employees)  &35  (23.2\%) \\
      Small size (11-50 employees)  &52 (34.4\%) \\
      Middle size (51-250 employees)  & 37 (24.5\%) \\
      Large size ($>$250 employees)  &27 (17.9\%) \\ \hline
      
      \textbf{Who performs test automation:} &  \\
      In-house test team  &95 (62.9\%) \\
      Outsourced test team &29 (19.2\%) \\
      Developers &76 (50.3\%) \\
      Other & 11 (7.3\%) \\ \hline
      
      \textbf{\% of automated test cases:} &  \\
      $<$10\%  & 31 (20.5\%) \\
      10-50\% &46 (30.5\%) \\
      50-90\% &33 (21.9\%) \\
      $>$90\% & 16 (10.5\%) \\ 
      We don't know & 8 (5.3\%) \\ 
      We don't measure & 17 (11.3\%) \\ \hline
  
     \textsuperscript{\textit{ab}} \textbf{Software development lifecycle:} &  \\
      \textit{\textbf{Agile}}: & \textit{\textbf{137}}  \textit{\textbf{90.7\%}}\\
      Scrum  &98 (64.9\%) \\
      Kanban &48 (31.8\%) \\
      Scaled Agile Framework  &25 (16.6\%) \\
      Feature Driven Development & 17   (11.3\%) \\ 
      Test-Driven Development & 16    (10.6\%) \\ 
      Behavior-Driven Development & 7   (4.6 \%) \\ 
      Lean Software & 11  (7.3\%) \\
      eXtreme Programming & 6 (4.0\%)\\ 
    \textit{\textbf{ DevOps}}& \textit{\textbf{43}}   (\textit{\textbf{28.5\%}})\\
     \textit{\textbf{Traditional}}: &\textit{\textbf{39}}  (\textit{\textbf{25.8\% }})\\
      Rational Unified Process& 5  (3.3\%) \\
      Waterfall or waterfall like &34 (22.5\%) \\
      \textit{\textbf{Other:}} & \textit{\textbf{6}}  (\textit{\textbf{ 4.0\% }})\\ 
      
      \bottomrule
    \end{tabular}
    
    \begin{tablenotes}
     \scriptsize \item [a] { Note that some organizational units are adopting several software development models.} 
     \scriptsize \item [b] { The classification of software development lifecycle models was noted in Section~\ref{sec:nbanalysis}}
   \end{tablenotes}
    \end{threeparttable}
  \end{center}
\end{table}

As for the percentage of automated cases, about half of the respondents reported that they automate less than 50\% test cases. Only about 10\% of respondents confirmed that they have over 90\% of automated test cases. 

With regard to the software development lifecycle, agile is almost  universally adopted (137 out of 151 responses). About one-third (N=43) of the responses reported that they use DevOps. Almost all (42 out of 43 responses) combine the adoption of DevOps with Agile models. For the traditional approach, one-fifth (N=34) of the responses stated that they follow a waterfall model, where 12 of them only follow this model. Several mentioned that Rational Unified Process is adopted.

\begin{figure}[ht]
  \centering
  \includegraphics[width=0.7\linewidth]{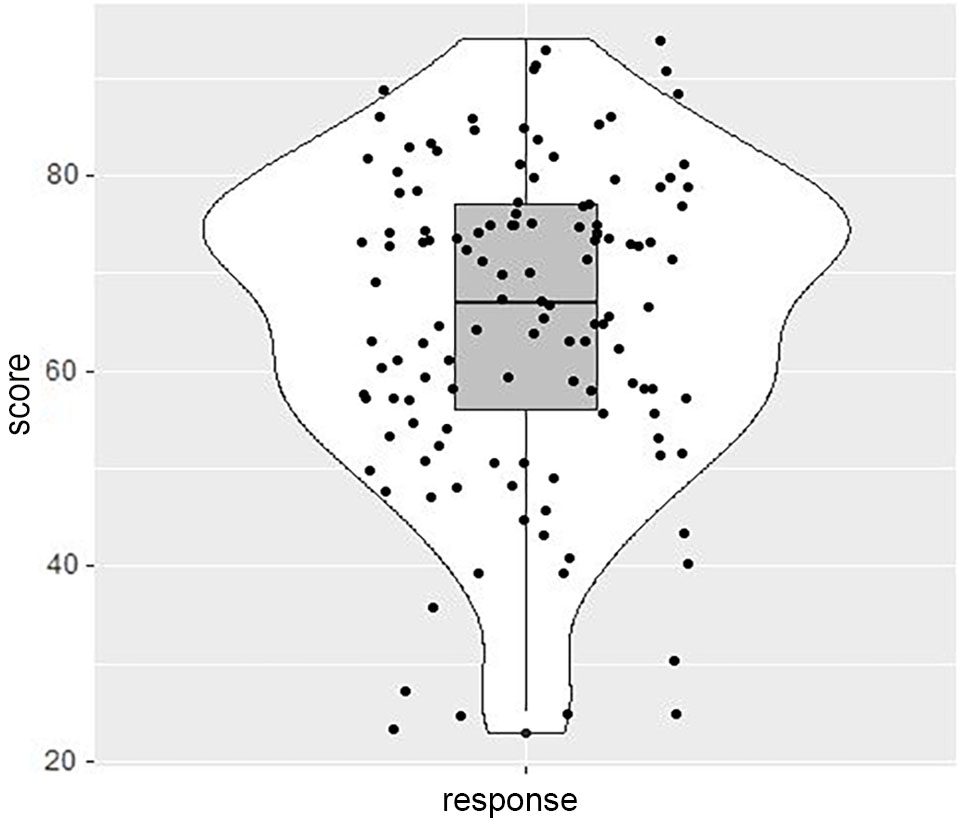}
  \caption{Overview of maturity scores of responses.}
  \label{fig:violinPlot1}
\end{figure}

\paragraph{\RQoneTitle.}
Figure~\ref{fig:violinPlot1} shows a violin plot with jittered dots\footnote{Violin plot - jitter https://www.tutorialgateway.org/r-ggplot2-jitter/} to visualize the distribution of total maturity scores of responses on test automation maturity questions in part 2 of our survey. One can see from this plot, the level of test automation maturity in different organizational units of respondents was differentiated by their maturity scores, i.e., from the Minimum(23) to the Maximum (94). 

The violin is skewed to the top, peaking around the interquartile range (IQR), aka. the midspread,  from 56 to 77. One can see that the range of IQR is still distant from both the Minimum (23) and Maximum (94). This implies that many test automation processes are relatively mature but there is the potential for the further improvement.

Several jittered dots above the score of 90 suggest that some response scores fairly close to the maximal points to be awarded (96) in the survey. This implies that some have reached a self-acclaimed high level of test automation maturity. In contrast, many dots are scattered in the first quartile, and there are several dots below the score of 30 in the violin. There is the long distance between those dots and the Maximum. These results imply that some are still very far from being mature, and there is plenty of room for improvement.

\begin{figure}[ht]
\centering
  \includegraphics[width=1\linewidth]{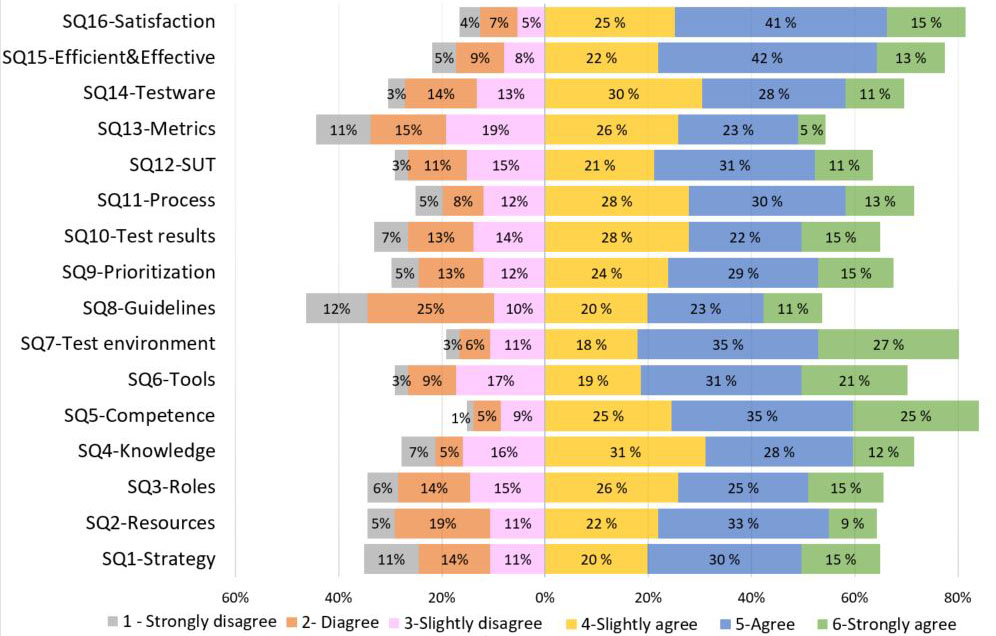}
  \caption{Survey responses on maturity questions}
  \label{fig:Surveyresponses}
\end{figure}

\paragraph{\RQtwoTitle.}
Figure~\ref{fig:Surveyresponses} shows the overview of responses to test automation maturity questions in part 2 of our survey. Agreed responses (i.e., who rate 4-6 from slightly agree to strongly agree) are stacked to the right of a vertical baseline on `0\%' on the x-axis. Disagreed responses (i.e., who rate 1-3 from strongly disagree to strongly agree) are stacked to the left of the same baseline. Note that since `no answer' exists for a certain survey question, the total percentage of all agreed responses and disagreed responses may be not equal to 100\%. 

The percentage of agreed responses on each question is in the range of 54\% (N=82) to
85\% (N=128). It may be noted that the situation is quite diverse with respect to different practices. We further compared the percentage of agreed responses and disagreed responses, in order to know which practices are more mature and immature in the industry: 

`\SQFive' has the largest percentage (85\%) of agreed responses, suggesting that, 85\% of responses agreed that their test team has enough expertise and technical skills for test automation. Besides, `\SQSeven' and `\SQSixteen' also have more than 80\% agreed responses. This indicates that the majority of responses agreed that they have the control over configuration of their test environment, and their automated tests can meet the given test purposes and bring substantial benefits.

`\SQEight' has the largest percentage (47\%) of disagreed responses. This suggests that there is lack of guidelines on designing and executing automated tests in general. In the free-text answers, 5 respondents highlighted the difficulty in defining such guidelines. It is followed by `\SQThirteen', which has 45\% disagreed responses. This means that many still do not have the right metrics to measure and improve test automation processes. 3 respondents indicated the lack of right metrics in the free text answers.

\paragraph{\RQthreeTitle.}

\begin{figure}[ht]
\centering
  \includegraphics[width=\linewidth]{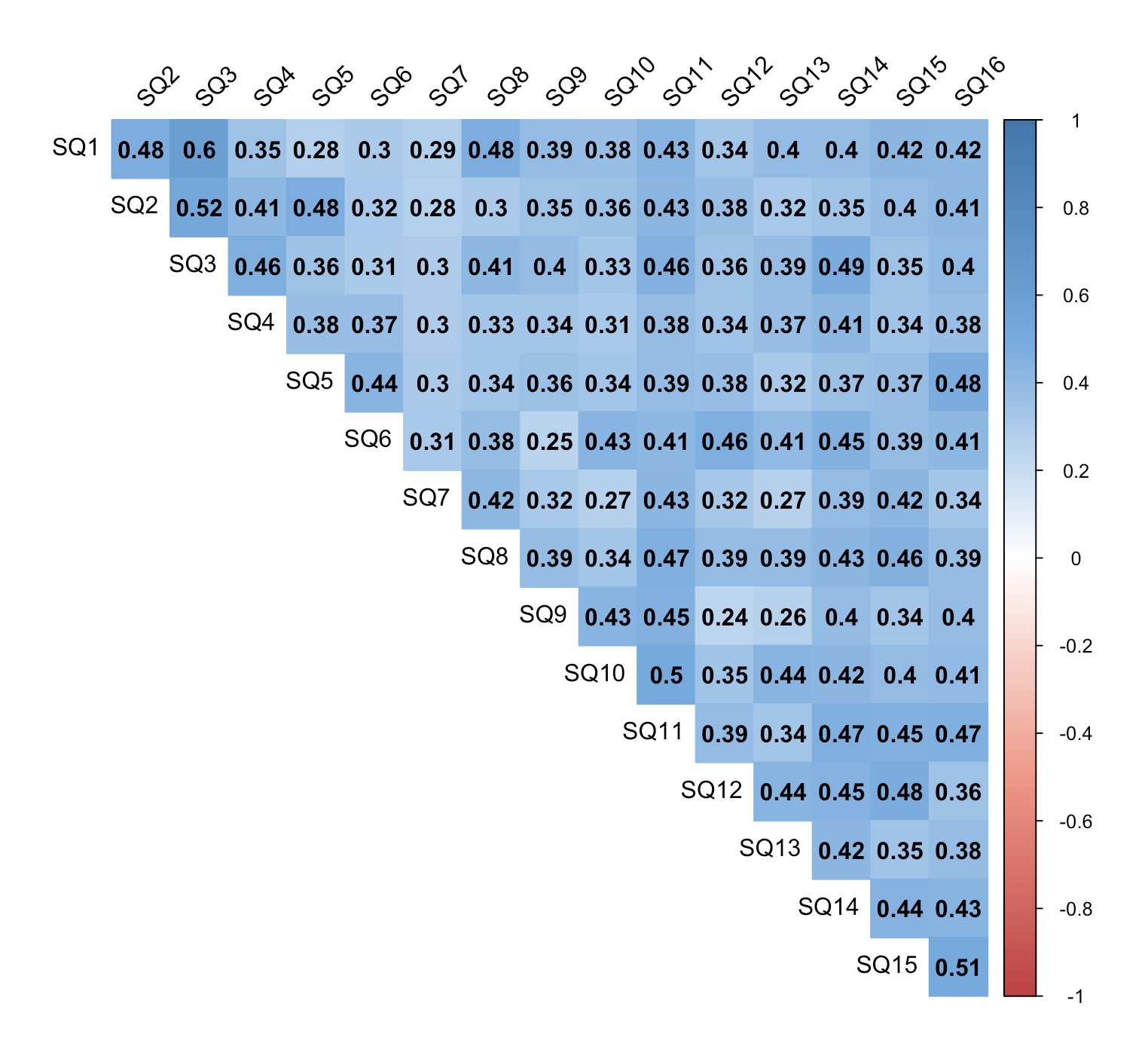}
  \vspace{-0.5em}
  \caption{Correlation matrix }
  \label{fig:corrplot1}
\end{figure}

Figure~\ref{fig:corrplot1} is a correlation plot that shows KR correlation coefficient $r_k$ values between each pair of test automation maturity questions in part 2 of our survey. As indicated by the color key, more negative values are represented by more dark red color and more positive values are represented by more dark blue. The number inside the color key represents the $r_k$. 

As shown in the correlation plot, $r_k$ ranges in values from 0.24 to 0.60. This indicates that adopted best practices are positively correlated with another in either the strong, moderate, or weak relationship. The number of pairs of survey questions having a moderate relationship ($r_k$=.30-.49) is largest. 


Table~\ref{tab:StrongCor} lists the pairs of survey questions that have a strong correlation ($r_k$ $\geqslant$ .50). PS01 has the highest $r_k$ = .60. This shows that the strongest correlation exists between the practices to define `test automation strategy’ and  `the roles and responsibilities of stakeholders'. 


  
  \begin{table}[ht]
  \begin{center}
  \scriptsize
    \caption{The strong correlation group}
    \label{tab:StrongCor}
    \begin{tabular}{ l p{5.2cm} l } 
    \toprule
      Ref. & The pair &\textbf{$r_k$}    \\
      \midrule
      PS01& `\SQOne'-`\SQThree' & .60 \\
      PS02& `\SQTwo`-`\SQThree' & .52 \\
      PS03& `\SQFifteen'-`\SQSixteen' & .51  \\
      PS04& `\SQTen'-`\SQEleven' & .50\\
      \bottomrule
    \end{tabular}
  \end{center}
\end{table}

\begin{figure}[ht]
\centering
  \includegraphics[width=0.8\linewidth]{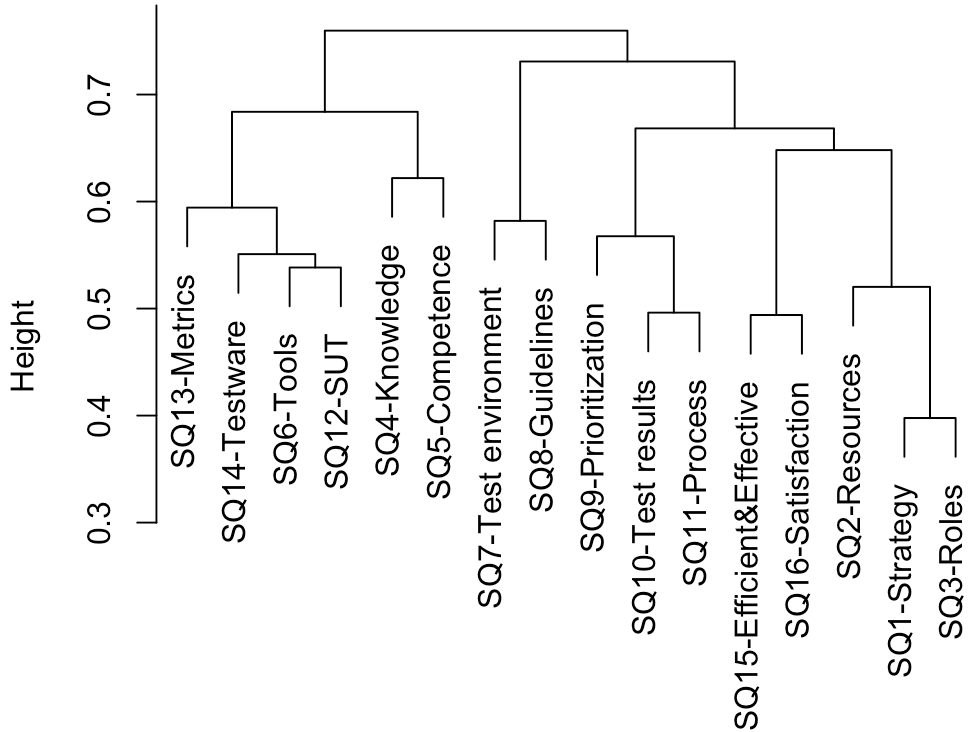}
  \caption{Cluster dendrogram}
  \label{fig:corrplot2}
\end{figure}

Figure \ref{fig:corrplot2}  is a  plot that shows the hirarchical clustering results based on $r_k$-based distance of test automation maturity questions in part 2 of our survey. Multiple test automation maturity questions clustered at lower Height (H) meaning that the practices (mentioned in those questions) have the close correlation. The number of clusters was identified to illustrate different types of practices that have the close correlation.

There is the close correlation among test management related practices. This can be seen from Figure~\ref{fig:corrplot2}, `\SQTwo' joints PS01 (`\SQOne'-`\SQThree') with H$\approx$.55. Besides, some technical related practices also are closely clustered. `\SQFourteen' joins the pair (`\SQSix'-`\SQTwelve') with H$\approx$.55. One can see that `\SQNine' joints the pair PS04 (`\SQTen'-`\SQEleven') at H$\approx$.55.



\paragraph{\RQfourTitle.}
Figure~\ref{fig:NB1} shows the NB regression model for RQ4. One can see some characteristics of organizational units are related to test automation maturity:

\begin{figure}[ht]
\centering
  \includegraphics[width=\linewidth]{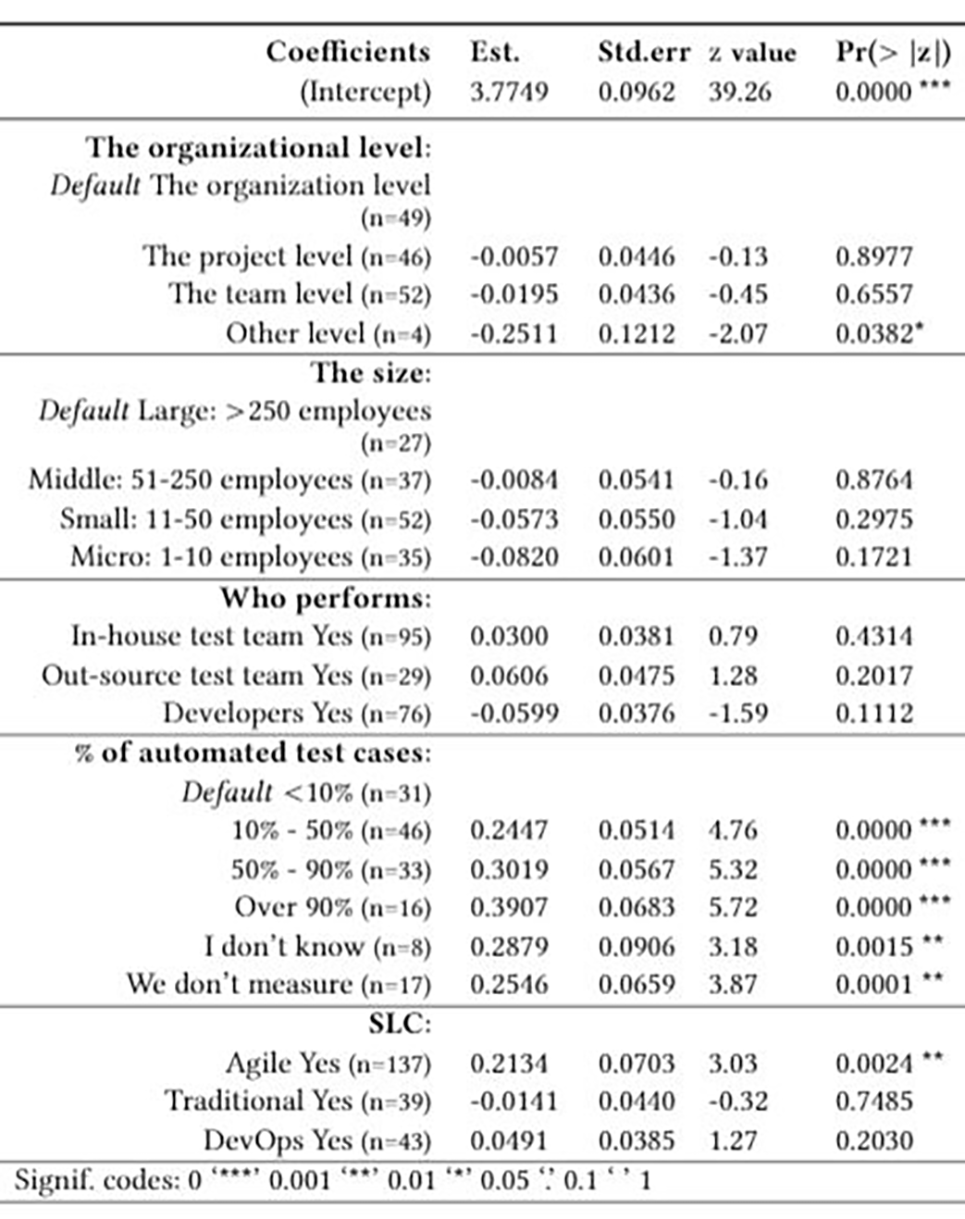}
  \caption{NB regression model 1}
  \label{fig:NB1}
\end{figure}

Test automation processes managed in the organization level tend to be associated with the higher level of maturity. This can be seen from Figure~\ref{fig:NB1}, `the organization level' was coded as the default, while other levels have a coefficient in the range of -0.0057 to -0.2511. However, the differences among those levels may be not statistically significant, since each level has  \textit{p-value $>$ 0.05.}



Larger organizational units appear to be associated with the higher level of test automation maturity, according to coefficients of the size variable. The default is `Large'.  `Middle', `Small', and `Micro' have a coefficient -~0.0084, -~0.0573, -~0.0820 respectively. 
The result is consistent with the predication for the organizational level variable, which indicates that test automation processes managed at the organization level tend to be more mature. However, as each of size variable has a \textit{p-value $>$ 0.05}, the differences among the sizes of organizational units are not statistically significant. 

Test automation processes performed by developers may be associated with the lower level of maturity. Regarding to `Who performs (test automaton)' related variables, `Developer Yes' is the only one has a negative coefficient -0.0599. `In-house test team Yes' and `Out-source test team Yes' have a positive coefficient 0.0300 and 0.0606 respectively. However, again the differences among `who performs (test automation)' options are not statistically significant, as the absence of significant codes in each variable.

The higher percentage of automate test cases is significantly associated with the higher level of test automaton maturity. `$<10\%$' was coded as the default. `10\%-50\%', `50\%-90\%', and `over 90\%' have a  coefficient 0.2447, 0.3019, and 0.3907 respectively. A coefficient of ‘10\%-50\%’ is less than a coefficient of 50\%-90\%’, which  is  less  than  a  coefficient  of‘over ‘90\%’. Nevertheless, all `10\%-50\%', 50\%-90\%', and `over `90\%' have a \textit{p-value = 0.0000}, suggesting the significant association. 

Test automation processes that follow the modern software development model Agile and DevOps tend to reach the higher level of maturity, compared to the ones following traditional Waterfall and Rational Unified Process models. `Agile Yes' and `DevOps Yes' have a positive coefficient 0.2134 and 0.0491 respectively, while `Traditional Yes' has a negative coefficient -0.0141. The presence of significant code `**' for `Agile Yes' confirms that the adoption of Agile methods is significantly associated with the higher level of test automation maturity. Besides, almost the all (42 out of 43) adopt the DevOps combined with Agile, as noted in Section~\ref{sec:organizationProfile}. This means that combining the adoption of DevOps and agile is associated with the higher level of test automation maturity than adopting agile alone.





\paragraph{\RQfiveTitle.}
 
\begin{figure}[ht]
\centering
  \includegraphics[width=\linewidth]{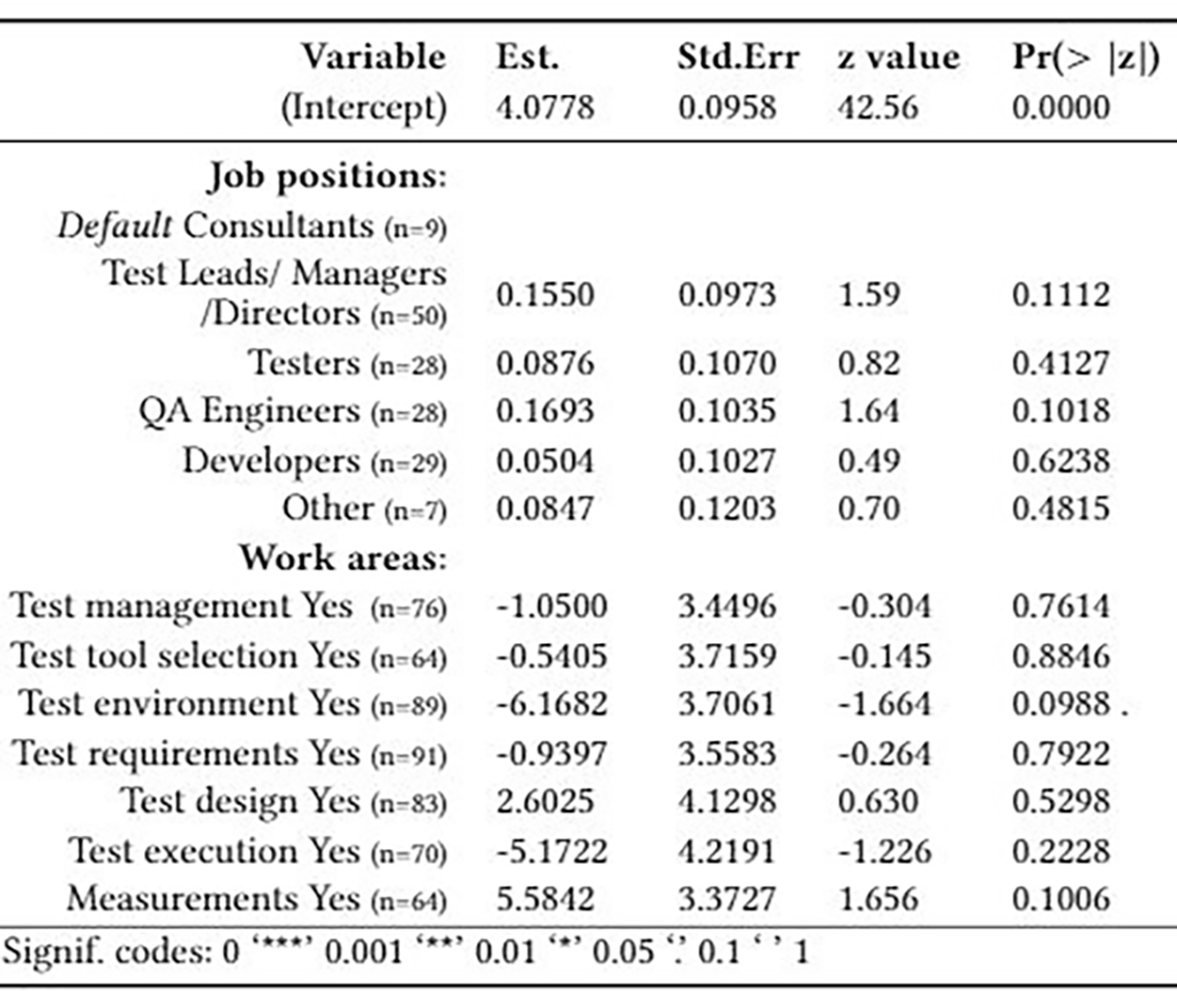}
  \caption{NB regression model 2}
  \label{fig:NB2}
\end{figure}

Figure~\ref{fig:NB2} is a NB regression model for `\RQfiveTitle'. It illustrates that the response variation among practitioners with different roles are explicable:
 

Consultants are likely to give the most pessimistic answers. Referring to the job position variable, one can see that compared to the default `Consultants', others have a positive coefficient in the range of 0.0876-0.1693.QA engineers tend to give the most optimistic answers, since `QA Engineers' has a highest-positive coefficient 0.1693. 

 Practitioners work in Test environment and Test execution KPAs provide the most pessimistic answers. The `Test environment Yes' variable has a lowest negative coefficient -6.1682. This is followed by `Test execution Yes' variable that has a coefficient -5.1722. The other work areas related variables have a coefficient greater than -1.0500, which is distant from them. 
 
 
The more optimistic answers are provided by practitioners working in Test design and Measurements KPAs. `Test design Yes' and `Measurements Yes' have a coefficient of 2.6025 and 5.5842,  while others have the negative one from -0.61682 to -0.5405. The variable `Measurements Yes' has the greatest positive coefficient, suggesting that practitioners work in the measurements KPA tend to give the most optimistic answers. 

The above results indicate that the current roles (regarding to job positions and work areas) of practitioners may lead to the response variation, but they are not statistically significant (p-value>0.05). 


\section{Discussion}
\label{sec:discussion}
\noindent We summarize survey results to each research question and discuss their implications.
\vspace{-5pt}
\paragraph{\RQoneTitle.}
\label{sec:dicussion_RQ1} Referring to the result of RQ1,  the level of test automation maturity in respondent organizations is differentiated by the practices they adopt. Some test automation processes are more mature than others based on adopted practices. This study result is aligned with the finding of ISTQB's recent Worldwide Software Testing Practices Report~\cite{ISTQB2018}, which surveyed thousands of test practitioners in the world. This report states that the level of test automation maturity may vary from one to another among organizations since test automation practices they performed are different. This indicates that there are potentials to further improve some test automaton processes, especially for the ones that are far from being mature, by performing recommended mature practices.

\paragraph{\RQtwoTitle.} 

 In our survey, most of responses agreed that their test team has enough expertise and technical skills for test automation, have the control over the configuration of their test environment, and they create automated tests to meet the given test purposes and bring substantial benefits. Those results are partially opposite with the World quality report 2018-19~\citep{worldqualityreport19}, which noted that the present industry performs immature practices to set up test environment and cultivate test automation expertise and skills of test teams. The possible reason could come from the differences in survey setup and respondents. Besides, based on our survey responses, there is lack of guidelines on 
  designing and executing automated tests and the right metrics to measure and improve test automation processes in general. We believe that those immature practices perceived by the practitioners are example gaps between academia and industry and need to be addressed. They point out important future research topics:
\begin{itemize}
    \item \textbf{Guidelines:} what guidelines should be provided to test practitioners for designing and executing automated tests, so that the development and maintenance effort will be minimal.
     \item \textbf{Test automation metrics:} what are the most important test automation metrics, how to find right test automation metrics, why the one thinks its current test automation metrics are not right. 
\end{itemize}

\paragraph{\RQthreeTitle.}\label{sec:dicussion_RQ2}~\cite{pocatilu2002} indicates that ``test automation practices are interdependence and consecutive, as each type of practice may result in intermediate deliveries to be used by others''. By extending his research, we confirmed that all test automation practices are positively correlated in practical context.  The description to some test automation practices in his research could explain some clusters of practices in our study :
\begin{itemize}
    \item A cluster of test management related practices (\SQOne-\SQThree-\SQTwo): Test automation strategy defines action steps that will be performed to allocate resources and define roles and responsibility of stakeholders. 
    \item A cluster of technical related practices (\SQSix-\SQTwelve-\SQFourteen): the use of suitable test tools may be helpful to create the maintainable testware and test automation testability features into a SUT.  
\end{itemize}

 Based on above discussion, we argue that, for practitioners, it is important to pay attention to the ripple effect of different practices, especially for the ones that are strongly correlated and closely clustered. Lack of any necessary practices may harm their test automaton maturity.

\paragraph{\RQfourTitle.}  We found that the high level of test automation maturity achieved in organizational units, which are managed at the organization level with large size, have in-house test team or/and out-source test team, automate the high percentage of test cases, and follow the modern software development model Agile and/or DevOps. Those results suggest that organization factors may affect test automation maturity. Some factors like percentage of automated cases and  the adoption of agile and DevOps have been examined by prior research~\citep{garousi2016}. We argue that the rest also should be studied. By identifying them and understanding their impact can help organizations to better use and implement test automation in the organizations.


\paragraph{\RQfiveTitle.} The results of RQ5 confirmed that the roles of respondents may lead to the response variation in our survey. The previous study~\citep{wang2019} pointed the response variation is likely to introduce the bias to assessment results of test automation maturity. Consequently, we believe that, when assessing test automation maturity of an organization, it is necessary to involve practitioners with different roles to avoid the assessment bias.  However, as our survey was anonymous, it is difficult to compare the impact of respondents' roles on the assessment result at a same organization. The further study is needed to validate what impact each type of roles has.

\section{Threats to validity}
\label{sec:threats}
\noindent The threats to this study and approaches taken to minimize their impact were explored, according to a standard checklist in software engineering from \cite{wohlin2012}. 

\textbf{Construct validity} refers to the extent to which the study present the theory
behind it. To ensure the construct validity,  we developed our survey according to the prior literature, reviewed it with industry experts, and did the small-scale piloting before actually executing it. The survey was designed and executed by following the standard guidelines in software engineering.





\textbf{Conclusion validity} is concerned with the extent to which correct conclusions are made through observations of the study. In this study, all the conclusions to each research question were drawn according to statistical results and are traceable to raw survey data. However, since this survey was anonymous and followed the GDPR, sharing the raw data of survey is not allowed.

\textbf{Internal validity} focus on how the study really cause the outcomes. In our study, threats to internal validity lie in the convenience sampling and survey execution.
The participants from diverse places are likely to bias the survey results. To avoid that, in the survey design, we studied main types of response bias and took the corresponding steps to control them, e.g., to avoid social desirability bias, the anonymity of the responses and result reporting were maintained. We measured the response quality and removed bad responses before analyzing and reporting survey results.


\textbf{External validity} is concerned with how the study results can be generalized. Selection bias may be a threat to external validity of this study. As most responses were received from Europe, population differences should be considered in the generalization of study results to rest of the world. 


\section{Conclusion}
\label{sec:conclusion}


\noindent Software organizations should assess and improve their test automation maturity for continuous improvement. They need a benchmark of the current state of their test automation processes and practices to identify improvement steps. 

In this paper, we conducted a test automation maturity survey with 151 practitioners coming from more than 101 organizations in 25 countries. Based on survey responses, we made several observations about the state of test automation practice in the present industry and discussed the implications of study results, see Section~\ref{sec:discussion}. 





This study has contributions to both academia and industry. It can help researchers and practitioners to understand the state of practice of test automation maturity in the present industry. For the industry side, the survey in this study and survey results together may help them to benchmark their test automation maturity and make the comparison with others  in the industry. This would help practitioners to better understand and conduct test automation processes. On the other side, our study is connected to the previous literature and extends the research in this area, as discussed in Section~\ref{sec:discussion}. By reviewing study results, researchers can find research topics which are interested to both academia and industry in the research scope of test automation maturity. Based on the findings we also suggested some follow-up research topics in this area, see Section~\ref{sec:discussion}. 





As a future study, we intend to do in-depth analysis of other factors that may affect test automation maturity, such as test tools and frameworks, test case design techniques, etc. We aim to integrate all results to establish a coherent framework for organizing current best practices in a validated improvement ladder.

\section*{\uppercase{Acknowledgements}}
\noindent This study is supported by TESTOMAT Project (ITEA3 ID number 16032), funded by Business Finland under Grant Decision ID 3192/31/2017.

\bibliographystyle{apalike}
{\small
\bibliography{ownbib}}

\end{document}